\documentstyle[12pt]{article}
\begin{document}
\def\p{\partial}
\def\na{\nabla}
\def\om{\omega}
\def\Om{\Omega}
\def\ga{\gamma}
\def\ph{\phi}
\def\ps{\psi}
\def\si{\sigma}
\def\la{\lambda}
\def\ep{\epsilon}
\def\ta{\tau}
\def\ka{\kappa}
\def\be{\beta}
\def\de{\delta}
\def\wh#1{{\widehat{#1}}}
\def\bskip{\baselineskip}
\def\cm{{\rm cm}\!\!}

\title{Beat-wave generation of plasmons in\\ semiconductor plasmas}
\author{V.I. Berezhiani\\
{\small\it International Centre for Theoretical Physics, Trieste, Italy} \\
{\small\it and}\\
{\small\it Institute of Physics, Tbilisi, Republic of Georgia}\\ \\
S.M. Mahajan\\
{\small\it Institute for Fusion Studies, The University of Texas at Austin,
Texas, USA}\\ {\small\it and}\\
{\small\it International Centre for Theoretical Physics, Trieste, Italy}}
\date{}
\maketitle
\bskip 22pt

\centerline{{\bf  Abstract}}
{{\bf  Address for correspondence}}:\\
V.I. Berezhiani \\
{\it International Centre for theoretical Physics},\\ P.O. Box 586, I-34100
Trieste, Italy.\\
FAX:+40 224163\\
E-MAIL: vazha@ictp.trieste.it
\clearpage
\bskip 24pt

Particle acceleration by the large-amplitude longitudinal waves excited in
an underdense
plasma by a propagating laser pulse (wake-field accelerator), or by the
beating of two
laser beams with nearly equal frequencies (beat-wave accelerator), is an
extremely
interesting and far-reaching idea [1]. The efforts to translate this
concept into reality,
however, have to surmount two serious problems: 1)~The creation of
relativistic plasmas
with $v^2_E/c^2\sim 1$ (necessary for these processes) requires enormous
field intensities
in excess of $10^{16}-10^{18}\,$W/cm$^2$, and 2)~the plasma has to be
extremely homogeneous [2]. These rather daunting requirements have made it
difficult even to carry out
exploratory experiments to test the proposed ideas.

Naturally, an alternative to standard plasma experiments, where these
severe constraints
could be mitigated, will be highly welcome. One could then begin initial
experimentation
to determine the validity of the theoretical framework, and shed light on
the eventual
feasibility of these ideas. Fortunately, such an alternative does exist; it
is provided by
certain special plasmas to be found in the narrow gap semi-conductors [3].
It just so happens
that in the two-band approximation of Kane's model dealing with narrow gap
semiconductors,
the Hamiltonian of the conduction band electrons mimics the relativistic
form $H=(m_*^2 c_*^2+c_*^2 p^2)^{1/2}$. Here $c_*=(E_g/2m_*)^{1/2}$ plays
the part of the speed of light, $m_*$ is the effective mass of the
electrons at the bottom of the conduction band, $E_g$ is the width of the
forbidden-band,
and $p$ is the electron quasimomentum.

This formal similarity would be just an interesting curio, but for the fact
that in several
materials, the characteristic velocity $c_*$ that enters in the Kane
dispersion law is much
less than the speed of light (for example $c_*\approx 3\cdot 10^{-3} c$ for
InSb). Because
of this, the jitter velocity of the electron fluid in the conduction band
can become
``relativistic" even when modest intensity electromagnetic fields are
applied. A relativistic jitter is possible because the conduction band is
partially empty, and as a
result the electrons can get accelerated under the effect of an electric
field. Nonparabolicity of the conduction band implies a nonlinear electron
velocity-momentum
dependence (${\bf v}=\p H/\p{\bf p}$) which, in turn, leads to a nonlinear
dependence of the
current density on the electric field. This nonlinearity dominates the
nonlinearity due to
electron heating provided the relevant wave frequencies are considerably
higher than the
effective collision frequency. Thus the dominant nonlinearity of this
system exactly
corresponds to the nonlinearity which lies at the foundation of the
acceleration schemes
mentioned earlier.

An expected consequence of this remarkable coincidence has been to use the methodologies of
relativistic plasmas to develop a pseudorelativistic dynamics for the
conduction electrons in order to delineate the optical properties of
narrow-gap semiconductors~[4]. The plasma physics of Kane-type
semiconductors has also been actively
investigated: in Ref.~[5], different kinds of parametric excitations of
density waves, and
parametric amplification of electromagnetic (EM) waves are presented, while
in Ref.~[6],
the authors explore the possibility of finding localized solitonic
structures, in addition
to studying the nonlinear self-interaction of EM waves in semiconductors.

In a recent publication~[7], we suggested that these semiconductor plasmas
could become a
veritable laboratory for testing the conceptual foundation of particle
accelerators based
on laser-plasma interaction. It was shown that an intense short laser pulse
propagating
through a semiconductor plasma generates a measurable longitudinal Langmuir
wave in its wake
(wake field) even for very moderate laser powers. Whenever the laser pulse
duration
$T_L\sim\om_e^{-1}$, the plasma frequency, wake-field excitation occurs.
This process is
naturally the analogue of the expected laser wake field excitation in the
usual relativistic plasma. This mechanism for the nonlinear coupling
between photons and plasmons
is an efficient way to produce finite-amplitude plasma excitations in
semiconductors with
readily available technology.

The next step in the investigation of `relativistic' semiconductor plasmas
is to explore if
the beat wave scheme for the generation of fast-moving large-amplitude
longitudinal waves
is also feasible. In this scheme, the plasma oscillation of frequency
$\om_e$ (the electron
plasma frequency) is resonantly excited by the ponderomotive force of two
propagating
collinear laser beams with nearly equal frequencies $\om_1$ and $\om_2$
($\om_1\sim\om_2$),
such that $\om_1-\om_2\sim\om_e\ll\om_1, \om_2$ (underdense plasma
condition). Because the
beat-wave generation of plasma waves is a resonant process, large-amplitude
plasma waves
can develop even for relatively weak lasers; the laser intensity
requirement could be
considerably less than what is necessary for the wake field generation
scheme. However,
experimental observation of beat wave excitation in normal gaseous plasmas
is difficult
because of the inherent inhomogeneities. Semiconductor plasmas, on the
other hand, are rather
homogeneous, and can be ideal for the experimental simulation of the
generic beat wave scheme. Since the nature of the induced longitudinal
field
will reflect important characteristics of the semiconductors, the
experiment could easily
lead to a new and exciting diagnostic as well. Although there are several
materials with a
nonlinear velocity-momentum relationship, we concentrate on semiconductors
of the Kane-type
because they, in addition, provide us with a laboratory to simulate (with
much smaller laser
intensities) a relativistic plasma.

For studying high frequency oscillations, the semiconductor plasma can be
treated as an
ideal electron fluid moving in a fixed ion lattice. The electrodynamics of
this system can, then,
be described by the set of electron fluid and Maxwell equations~[4-7]:

$${{\bf \na}}\times{\bf B}={\ep_o\over c}\,{\p{\bf E}\over\p t}- {4\pi e\over c}\,
n{\bf v}\eqno(1)$$
$${{\bf \na}}\times{\bf E}=-{1\over c}\,{\p{\bf B}\over\p t}\eqno(2)$$
$$\ep_o{{\bf \na}}\cdot{\bf E}=4\pi e(n_o-n)\eqno(3)$$ $${\p n\over\p
t}+{{\bf \na}}\cdot(n{\bf v})=0\eqno(4)$$ $${\p{\bf p}\over\p
t}+({\bf v}\cdot{{\bf \na}}){\bf p}=-e{\bf E}-{e\over c}\,({\bf v}\times{\bf B})\eqno(5)$$
where $n$ is the electron concentration, $\ep_o$ is the dielectric constant
of the lattice and ${\bf p}$ is the quasimomentum of the conduction-band
electrons. This system is augmented by the nonlinear velocity-momentum
relation $${\bf v}={{\bf p}\over m_*(1+p^2/m_*^2c_*^2)^{1/2}}\ .\eqno(6)$$
Equation~(6) is valid if $\om\gg\nu$, where $\om$ is the characteristic
wave frequency and $\nu$ is the electron effective collision frequency.
Using the condition for the absence of generalized vorticity $${\bf B}={c\over
e}\,{{\bf \na}}\times {\bf p}\eqno(7)$$ and Eq.~(6), Eq.~(5) can be rewritten in
the simplified form $${\p{\bf p}\over\p
t}+m_*c_*^2{{\bf \na}}\ga=-e{\bf E}\eqno(8)$$ where $\ga$ is the `relativistic'
factor $$\ga=(1+p^2/m_*^2c_*^2)^{1/2}.\eqno(9)$$

For simplicity we consider a one-dimensional problem, the laser radiation
propagates along the $z$ axis, and all physical quantities depend on only
the space coordinate $z$, and time $t$. For this case, the transverse
component of
the momentum equation~(8) can be written as: $${\p{\bf p}_\bot\over\p
t}=-e{\bf E}_\bot\eqno(10)$$ which, on integration, yields
${\bf p}_\bot=(e/c){\bf A}_\bot$ where ${\bf A}_\bot$ is the vector potential of the
electromagnetic (EM) field. The longitudinal
electron motion obeys
$${\p p_z\over\p t}+m_* c_*^2{\p\over\p z} \left(1+{(p_\bot^2+p_z^2)\over
m_*^2c_*^2}\right)^{1/2}=e{\p\varphi\over\p z}\eqno(11)$$
where $\varphi$ is the scalar potential associated with the excited
longitudinal field.

It is evident from Eq.~(11) that the laser radiation, represented by the
$p^2_\perp$ term, always
generates a longitudinal response. The details of the response will surely
depend upon the nature
of the inducing laser radiation. In order to investigate the resonant
processes inspired
by the beat-wave concept, we begin with two linearly polarized beams of
frequencies
$\om_1$ and $\om_2$ with $\om_1-\om_2\sim\om_e\ll\om_1, \om_2$. The
transverse component of the induced electron momentum can then be written
as (${\bf p}_\bot\sim A_\bot$),
$${\bf p}_\bot=\wh{{\bf x}} p_1\exp(-\om_1t+k_1 z)+ \wh{{\bf x}} p_2\exp(-\om_2
t+k_2 z)+ \rm c.c.\eqno(12)$$ where $k_{1,2}$ are the mean wave numbers
associated with the laser beams, and the amplitudes
$p_1, p_2$ are related to the laser electric fields by
$p_{1,2}=ieE_{1,2}/\om_{1,2}$.

Note that the transverse waves satisfy the linear dispersion relation
$\om_{1,2}^2=\om_{*e}^2+\ep_o^{-1} k_{1,2}^2 c^2$, where $\om_{*e}=(4 \pi
e^2n_o/\ep_om_*)^{1/2}$ is the effective Langmuir frequency. If
$\om_{1,2}\gg\om_{*e}$ then the phase velocity of the plasma wave
$\om_{*e}/(k_1-k_2)$ is equal to
$v_g=c\ep_o^{-1/2}(1-\om_{*e}^2/\om_{1,2}^2)^{1/2}$, the group velocity of
the laser beam. Since $c\ep_0^{-1/2}$ is the speed of light ($\sim$ phase
velocity of the lasers) in the semiconductor, all relevant phase speeds are
nearly equal. We can
then safely assume that the resultant wave motion is a function of the
single variable
$\ta=t-z/v_g$, where
$v_g\approx c\ep_o^{-1/2}$. The transparent (underdense) plasma
($\om_{1,2}\gg\om_{*e}$)
assumption offers an additional simplification; we may assume that, during
the interaction time of
interest, the laser fields ($p_{1,2}, E_{1,2}$) remain unchanged and can be
presumed to be
constant~[7].

Using Eqs.~(3), (4), and (11) and choosing the initial conditions that
there is no longitudinal motion at $\ta=0$, the scalar potential of excited
longitudinal field $\varphi$ is found to satisfy

$${d^2\ph\over d\ta^2}=\om_{*e}^2 {u_g^2\over(1-u_g^2)}
\left[{u_g(1+\ph)\over[(1+\ph)^2-(1-u_g^2)(1+p_\bot^2)]^{1/2}}-1\right]\eqno
(13)$$
where we have introduced the new dimensionless variables ${\bf p}={\bf p}/m_*
c_*$, $\ph=e\varphi/m_* c_*^2$ and $u_g\approx (c/{c_{*}} \ep_o^{1/2})$.
Equation~(13) is very similar to
the equation for the wake field $\varphi$ driven by a single laser pulse~[7].

In order to demonstrate that $\varphi$ of Eq.~(13) can be resonantly
driven, let us
assume $p_\perp^2\ (p_\perp\sim A_\perp)\ll 1$. This, along with the fact
that $u_g\left(\sim c/c_*\ep^{1/2}, c\gg c_{*}\right)\gg 1$, converts
(13) to $\left(\ep=u^{-2}_g\ll 1\right)$
$${d^2\ph\over d\ta^2}+\om_{*e}^2\ph-\ep{\om_{*e}^2\over
2}\,(3\ph^2+\ph^3)=\om_{*e}^2 {p_\bot^2\over 2},\eqno(14)$$ an equation
representing a driven nonlinear oscillator.

The driving force $\sim p_\bot^2$ (right-hand side of Eq.~(14)) contains a
term with frequency $\Om=\om_1-\om_2$ and if $\Om\approx\om_{*e}$, the
plasma wave can be
resonantly excited. The nonlinear terms on the left-hand side of~(14) are
due to the ``relativistic" self nonlinearity of the longitudinal
motion, and follow from the nonparabolicity of the conduction band
dispersion relation. If we neglect the nonlinear terms in Eq.~(14) (i.e.
$\ep\to 0$), the resonant mode will have secular growth. {\em The
relativistic nonlinearity provides a
saturation mechanism} without which the model will be quite unphysical. By
employing simple analytical tools, we can extract much information from
Eq.~(14). Following standard procedure, let
us seek a solution of the form~[8],
$$\ph=\ph_o(\ta)\exp(-i\Om\ta)+ \rm c.c.\eqno(15)$$ where $\ph_o(\ta)$ is a
slow time dependent amplitude. Clearly Eq.~(15) can be an approximate
solution only when we concentrate only on the resonant terms. After simple
manipulations, the
resonant terms lead to the leading order equation $$i{d\ph_o\over
d\ta}+\de\cdot\ph_o+\be\vert\ph_o\vert^2\ph_o=-\la,\eqno(16)$$ where
$\de=\Om-\om_{*e}$ is the frequency mismatch, $\be=(3/4)\ep\om_{*e}$ and
$\la=(1/2)\om_{*e} p_1\cdot p_2^*$. To proceed further, we let
$\ph_o=A\,\exp(i\ps)$, and obtain
two coupled first order equations for $A$ and $\ps$ [8]. Using the obvious
initial condition
$A(0)=0$, and eliminating $\ps$, we find $$\left({dA\over
d\ta}\right)^2=\la^2-\left({\be\over 4}A^3+{\de\over
2}\,A\right)^2,\eqno(17)$$
from which, among other things, the steady state amplitude $(dA/d\ta=0)$
can be easily inferred.
For perfect phase matching ($\de=0$), the saturation amplitude is
$$A_{max}=\left({4\la\over\be}\right)^{1/3}=\left({8\over 3\ep} \vert
p_1\vert\cdot\vert p_2\vert\right)^{1/3},\eqno(18)$$ and the initial growth
is linear in time, $A=\la\ta=0.5\omega_{*e}\vert p_1\vert\vert
p_2\vert\ta$. It is also straightforward to estimate the saturation time,
$$\ta_{\rm sat}\approx
2.8\om_{*e}^{-1}\ep^{-1/3}(|p_1|\cdot|p_2|)^{-2/3}.\eqno(19)$$

For a finite mismatch ($\de\ne 0$), the saturation amplitude $A_\de$ can be
obtained solving the cubic (right-hand side of Eq.~(17)),
$$A_\de^3+{2\de\over\be}\, A_\de-{4\la\over\be}=0.\eqno(20)$$ Notice that
due to the nonlinear term $\be|\varphi_0|^2$ in Eq.~(16), the maximum
amplitude does
not correspond to $\de=0$. Analysis of Eq.~(20) shows that the maximum
value of $A_{\de max}=1.6
A_{max}$ corresponds to the frequency mismatch
$$\de=-1.3\om_{*e}\ep^{1/3}\, (|p_1|\cdot|p_2|)^{2/3}.\eqno(21)$$

Approximate analytical formulas (18)--(21) can help us gauge the efficiency
of the beat wave
mechanism. Let us take the $n$-InSb plasma for which the required
parameters are: $T=77\,$K,
$m_*=m_e/74$, $\ep_o=16$, and $c_*=c/253$. For the $CO_2$ laser beams with
respective vacuum wave
lengths $\la_1=10.81\mu\,$m and $\la_2=10.6\mu\,$m, the resonant Langmuir
frequency is
$\om_{*e}=3.5\cdot 10^{12}$ corresponding to a plasma density $n=8\cdot
10^{14}\cm^{-3}$, and plasmon wave length $\la_e=0.13\,$mm. For these
parameters, the electron
collision frequency $\nu\sim 10^{-2}\om_{*e}$. Assuming that the two laser
beams have equal intensities with an electric field $E_{1,2}=10^5 V/\cm$,
or equivalently,
$p_\bot^2\approx 0.4$, the maximum longitudinal field can grow to
$E_l\approx 120 V/$cm in a
total time $\ta_{\rm sat}\approx 80\om_{*e}^{-1}$. Note that the saturation
time [Eq.~(19)] is
larger for smaller amplitudes of laser radiation. When the time for
reaching saturation is
greater than the collision period, the collisional effects must be
included. This could be
achieved by formally adding the term
$\nu\p\ph/\p\ta$ on the left-hand side of Eq.~(14). We do not consider this
effect here, it
will trivially reduce the amplitude of the longitudinal field, in addition
to introducing
bistability and hysteresis.

We have thus demonstrated that strong, easily measurable longitudinal waves
can be excited by two
near frequency laser beams propagating in an $n$-type InSb plasma. The
character of the excited
wave, of course, depends not only upon the defining parameters of the
system, but also on the
details of the beat-wave mechanism. One of the very important aspects of
the proposed beat-wave and
similar schemes is that the relativistic nonlinearities will provide a
saturation mechanism. This
general hypothesis, among others, can be readily tested in a cheap, readily
available,
`relativistic,' and highly uniform semiconductor plasma. Experimentation of
this type could
answer several preliminary questions as to the eventual viability of these
schemes, and provide
a basic diagnostic for the nonlinear optical properties of solid state
systems to boot.

This work was supported, in part, by the U.S. department of Energy Contract
No. DE-FG05-80ET-53088.

\newpage

\centerline{REFERENCES}
\begin{description}
\item{[1]} T. Tajima and J.M. Dawson, Phys. Rev. Lett. {\bf 43} (1979)
267;\\ P. Sprangle, E. Esary, A. Ting, and G. Joyce, Appl. Phys. Lett.
{\bf 53} (1988) 2146;\\
J.M. Dawson, Physica Scripta~T {\bf 52} (1994) 7. \item{[2]} R.J. Noble,
Phys. Rev. A {\bf  32} (1985) 460;\\ W. Horton and T. Tajima, Phys. Rev. A
{\bf  31} (1985) 3937;\\ L.M. Gorbunov and V.I. Kirsanov, Sov. Phys. JETP
{\bf  69} (1989) 329;\\ G. Matthieussent, Physica Scripta T {\bf  50} (1994)
51. \item{[3]} O. Kane, J. Phys. Chem. Solid {\bf 1} (1957) 249. \item{[4]}
N. Tzoar and J.I. Gersten, Phys. Rev. Lett. {\bf 26} (1971) 1634;\\ N. Tzoar
and J.I. Gersten, Phys. Rev.~B {\bf 4} (1971) 3540. \item{[5]} P.A. Wolff
and G.A. Pearson, Phys. Rev. Lett. {\bf  17} (1966) 1015;\\ J.I. Gersten and
N. Tzoar, Phys. Rev. Lett. {\bf 27} (1971) 1650;\\ F.G. Bass and V.A.
Pogrebnyak, Sov. Phys. Solid State {\bf 14} (1972) 1518;\\ L.A. Ostrovskii
and V.G. Yakhno, Sov. Phys. Solid State {\bf 15} (1973) 306;\\ Ya.I.
Kishenko and N.Ya. Kotsarenko, Sov. Phys. Solid State {\bf 18} (1976)
1920;\\ V.S. Paverman and N.A. Papuashvili, Sov. Phys. Semicond. {\bf 19}
(1985) 852. \item{[6]} K.A. Gorshkov, V.A. Kozlov, and L.A. Ostrovskii,
Sov. Phys. JETP {\bf 38} (1974) 93;\\
F.G. Bass, L.B. Vatova, and Yu.G. Gurevich, Sov. Phys. Solid State {\bf 15}
(1974) 2033;\\
F.G. Bass and S.I. Khankina, Sov. Phys. Semicond. {\bf 18} (1984) 220.
\item{[7]} V.I. Berezhiani and S.M. Mahajan, Phys. Rev. Lett. {\bf 73}
(1994) 1837. \item{[8]} M.N. Rosenbluth and C.S. Liu, Phys. Rev. Lett.
{\bf 29} (1972) 701;\\ C.M. Tang, P. Sprangle, and R.N. Sudan, Phys. Fluids
{\bf 28} (1985) 1974;\\ R. Bingam, R.A. Cairns, and R.G. Evans, Plasma Phys.
and Controlled Fusion, {\bf 28} (1986) 1735.

\end{description}

\end{document}